\documentclass[3p,a4paper]{elsarticle}
\usepackage{amsmath}
\usepackage{amssymb}
\usepackage{graphicx}
\usepackage{epsfig}

\begin{document}

\newcommand{\tpi}{2 \pi \times}	
\newcommand{\half}{\frac{1}{2}}
\newcommand{\vc}{\mathbf}

\newcommand{\ra}{\rangle}
\newcommand{\la}{\langle}
\newcommand{\ket}[1]{|#1\rangle}
\newcommand{\bra}[1]{\langle #1|}

\newcommand{\muB}{\mu_\mathrm{B}}
\newcommand{\Jop}{\hat{\mathbf{J}}}
\newcommand{\Iop}{\hat{\mathbf{I}}}
\newcommand{\ephat}{\mbox{\boldmath$\hat{\epsilon}$}}

\newcommand{\hel}{\mathcal{H}_{\mbox{\scriptsize el}}}
\newcommand{\hfs}{\mathcal{H}_{\mbox{\scriptsize FS}}}
\newcommand{\hhfs}{\mathcal{H}_{\mbox{\scriptsize HFS}}}
\newcommand{\hmd}{\mathcal{H}_{\mbox{\scriptsize B}}}
\newcommand{\hamio}{\mathcal{H}_0}

\begin{frontmatter}
\title{Experimental methods of ultracold atomic physics}
\author[berkeley,lbl]{D.M. Stamper-Kurn}
\author[toronto,cifar]{J.H. Thywissen}

\address[berkeley]{Department of Physics, University of California, Berkeley CA 94720}
\address[lbl]{Materials Sciences Division, Lawrence Berkeley National Laboratory, Berkeley, CA 94720}
\address[toronto]{Department of Physics, University of Toronto, Ontario, M5S 1A7 Canada}
\address[cifar]{Canadian Institute for Advanced Research, Toronto, Ontario, M5G 1Z8 Canada}

\begin{abstract}
Experiments on solid-state materials and atomic quantum gases are increasingly investigating similar concepts in many-body quantum physics.  Yet, the flavor of experiments on the gaseous atomic materials is different from that of conventional materials research.  Here, we summarize some aspects of atomic physics and some of the common technical elements of cold-atom experiments which underlie the investigations described in the remaining chapters of this volume.
\end{abstract}

\date{\today}
\end{frontmatter}

\tableofcontents

\clearpage

The broad appeal of research on quantum gases relies on the universality of many-body quantum physics.  For example, regardless of whether it is constructed using electrons (as in superconductors), neutrons (as in neutron stars), or neutral atoms of different hyperfine spin (as in ultracold lithium gases), a system of strongly interacting, mobile fermions will show the same phenomenology.  Similarly, nonlinear and quantum optics can be treated in a common theoretical scheme, regardless of whether the bosonic fields under investigation correspond to massless photons or massive rubidium atoms.  Such universality allows ultracold atomic physics to contribute significantly to fields as diverse as condensed matter physics, high energy astrophysics, and quantum optics.  As such, the collection of works within this book is meant to bridge the gap between practitioners of these diverse fields so as to make the exchange among them more productive.

Yet, despite these appealing similarities, there do remain system-specific considerations that must be kept in mind in comparing physical systems built from different basic ingredients.  This chapter discusses some of the atom-specific aspects of ultracold atomic physics experiments.  We focus on two main topics: the common experimental techniques of quantum gas experiments and the nature of atom-atom interactions.  Our treatment will be at a colloquium level; for more detailed explanations, references are given.

\section{Introduction: Why so cold?}
\label{sec:intro}

There is little variety in temperature and density among the quantum-degenerate neutral gases produced currently in over one hundred laboratories in seventeen countries.  This commonality may be surprising because, unlike solids, gases have no lower bound in density. However to within a factor of ten, the density of an ultracold gas is $n \approx 10^{13}$\,cm$^{-3}$, about six orders of magnitude lower than the density of an ideal gas at standard temperature and pressure. Why this particular density? For $n \gtrsim 10^{15}$\,cm$^{-3}$, loss processes, such as a three-particle collision leading to the formation of a deeply bound molecule, become faster than rethermalization from elastic collisions. For $n \lesssim 10^{12}$\,cm$^{-3}$, the characteristic energy and temperature scales of the quantum gas become impractically small, particularly if one is interested in anything beyond the non-interacting ideal gas, and the thermalization rate of the gas becomes slow compared to parasitic heating rates and the vacuum-limited lifetimes of the frigid gas samples.

This common density sets a common characteristic energy and length scale for experiments on cold gases. From the inter-particle spacing $n^{-1/3}$, typically 300~nm, one would guess the energy scale for the physics of such gases to be $h^2 n^{2/3}/ M$, where $h=2 \pi \hbar$ is Planck's constant and $M$ is the atomic mass; more formally, one often takes the Fermi energy $E_F = \hbar^2 k_F^2/ 2 M$, where $k_F = (6 \pi^2 n)^{1/3}$ is the Fermi wave vector. For $^{87}$Rb, this energy scale is around $4 \times 10^{-11}$\,eV, corresponding to a temperature of about 500~nK, and a frequency of about 10~kHz.\footnote{Masses of degenerate neutral atoms range from 1 to 174 atomic units, so we have taken $^{87}$Rb as a typical example. Rubidium was also the first gaseous element to be Bose condensed, and is still the most commonly used species for ultracold boson experiments.}%

This energy scale also (roughly) defines the temperature at which a gas becomes quantum degenerate.  The onset of quantum degeneracy can also be considered from the comparison of length scales.  From the system temperature, we can define the thermal deBroglie wavelength,
\begin{equation}
\lambda_T = \sqrt{\frac{2 \pi \hbar^2}{M k_B T}},
\end{equation}
where $k_B$ is the Boltzmann constant.  When the de Broglie wavelength is comparable to the interparticle spacing, the coherent matter waves associated with the various particles in the gas are forced to overlap, meaning (pictorially) that the number of independent quantum states in the gas becomes comparable to the number of gas particles.  At this point, the quantum statistics of particles come into play in describing the nature of the gas.  The ultra-low temperature of quantum gases is therefore simply a consequence of their necessarily low density.  Since the gases in question are roughly a billion times less dense than liquid helium, they are degenerate at a temperature a million times lower than the lambda point of helium.

The low temperature scale required for the study of quantum degenerate gases has been, and continues to be, the prime technical challenge in ultracold-atom research.  As discussed in Sec.~\ref{sec:cycle}, cooling gases from room temperature to quantum degeneracy relies on a hybrid of cooling methods. These methods took decades to develop. Even today, with cutting-edge techniques, gases cannot match the extreme quantum degeneracy of electrons within a solid in a dilution refrigerator, where the ratio $k_B T / E_F$ is around  $10^{-6}$; in contrast, for cold atom experiments to date, this ratio goes no lower than $10^{-2}$.  Thus, advancing the frontier cold atom experiments still requires the continued development of cooling techniques. Pursuing this frontier is tremendously appealing, as one expects that neutral gases would be able to explore the rich physics of spin liquids, topological quantum matter, pure BCS superfluidity, and perhaps d-wave pairing in lattices \cite{stam09shift}.

However, even within the current technical limits, the combination of low density and ultracold temperature creates an extraordinary opportunity to study many-body systems with cold atoms.  Atomic gases are subject almost exclusively to pair-wise interactions that can be characterized completely by considering the physics of an isolated two-atom molecule.  Now, a molecule of, say, two $^{87}$Rb atoms is not the simplest quantum object, and its properties cannot be determined \emph{ab initio}.  However, after taking into account a range of data from various forms of molecular spectroscopy, it is possible to set up a reliable model that describes the scattering properties of two isolated atoms extremely well, and, thereby, provides the only inputs needed to specify an accurate description of a many-atom quantum gas comprising $^{87}$Rb atoms.  Moreover, at the very low kinetic energies characteristic of atomic gases, the partial-wave treatment of binary collisions is of great utility in reducing the complete characterization of atomic interactions simply to the specifications of the s-wave scattering length, $a_s$, and, on rare occasions, to the range of the potential, $r_0$. As explained in Sec.~\ref{sec:interactions}, the scattering length can be controlled by tuning the structure of the two-atom molecular potential, e.g.\ using magnetic-field-tunable Feshbach resonances.

For such reasons, ultracold-atom ``materials'' can be regarded as faithful experimental renditions of a known many-body Hamiltonian. Experiments on quantum gases have therefore been dubbed as \emph{quantum simulations} of various models (i.e.\ Hamiltonians) of many-body quantum physics.  Although cold-atom systems have non-universal eccentricities just like any other materials, it is comforting to know that even these eccentricities can, in principle, be characterized \emph{ab initio}.

\section{Manipulation of atoms and molecules}
\label{sec:AMO}

In this section we hope to provide the reader with ``(almost) everything you wanted to know about a neutral atom but were afraid to ask.'' Mostly, we will concern ourselves with the magnetic and electric susceptibility of a single ground state atom, the latter due to coupling of the ground state to excited states by electric fields. In Sec.~\ref{sec:interactions} we will extend this discussion to include interactions between two ground-state atoms.

\subsection{Atomic structure basics}
\label{sec:structure}

At the time of writing this chapter, eleven elements have been Bose-condensed: hydrogen (H), five alkali metals (Li, Na, two isotopes of K, Rb, Cs), three earth-alkaline and similar metals (Ca, three isotopes of Sr, four isotopes of Yb),  two transition metals (Cr, Dy), and one metastable noble gas (He*).
Five of these elements also have naturally occuring fermionic isotopes that have been cooled to quantum degeneracy: $^3$He*, $^6$Li, $^{40}$K, $^{87}$Sr, and $^{173}$Yb.
Alkalis were the first to be Bose-condensed, and remain popular due to the simplicity of their electronic structure: a single unpaired electron is optically active, just as in hydrogen, while the remaining electronics form a relatively inert core.

In the absence of external fields, there are three terms in the Hamiltonian of hydrogen-like atoms,
\begin{align*}
\hamio=\hel+\hfs+\hhfs,
\end{align*}
where $\hel$ contains the non-relativistic kinetic energy of the electrons and the Coulomb interaction between them; $\hfs$ is the ``fine structure'' term that includes relativistic corrections to $\hel$, electron spin, and spin-orbit terms; and $\hhfs$ is the ``hyperfine structure'' term induced by the spin and electric quadrupole moment of the nucleus.

In the Russell-Saunders ($L S$) coupling scheme, which is relevant when the electronic valence shells are sparsely populated as is the case for hydrogen-like atoms, the eigenstates of $\hel + \hfs$ share a common set of quantum numbers: $n$ (principal quantum number), $L$ (total orbital angular momentum), and $J$ (total overall angular momentum). For instance, the ground state of Rb is $5^2\mathrm{S}_{1/2}$, meaning the optically active electron is in the $n=5$ orbital, with $L=0$, and an angular momentum of $J=1/2$. The ``2'' superscript gives the degeneracy for the total electronic spin of $1/2$.

For alkali atoms, the two lowest-lying excited states are $n^2\mathrm{P}_{1/2}$ and $n^2\mathrm{P}_{3/2}$.  Different from the situation in hydrogen, the energies of these levels differ from that of the $n$S state because the higher angular momentum of the P states pulls the electron orbit to a higher radius, sampling less of the charge density of the core electrons. Such a many-electron mechanism (known as the ``quantum defect'') is weak for excited states with higher principal quantum numbers. For the alkalis, this $n$S-$n$P transition is the lowest-energy and strongest optical transition. The resonant frequency occurs at several hundred THz (roughly 1.5 eV of energy), and is therefore accessed by light in the near-infrared to visible range, generated easily from solid-state or dye laser systems.

For atoms with nuclear spin, $\hhfs$ further splits the fine-structure eigenstates. The total angular momentum, which is a sum of the electron orbital and spin angular momentum and of the nuclear spin, is called $F$. For example, for $^{87}$Rb the nuclear spin is $I=3/2$, which, added to the ground-state electronic angular momentum of $1/2$ (all of it from spin), gives either $F=1$ or $F=2$. These states are split by 6.8~GHz in $^{87}$Rb, and similar amounts in other hydrogen-like atoms, small enough that the upper state does not decay spontaneously.

\begin{figure}[th!]
\begin{center}
\includegraphics[width=10cm]{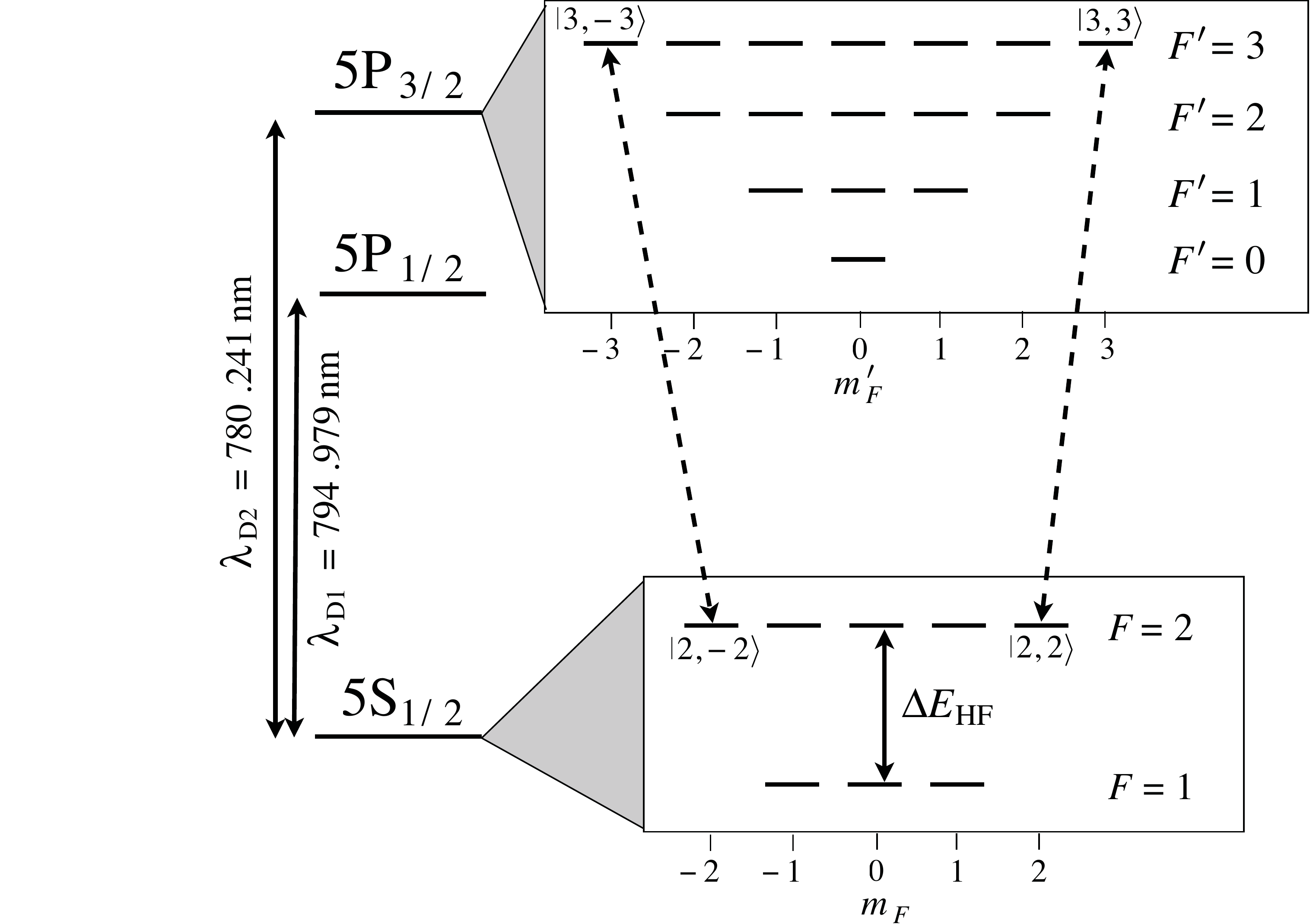}
\end{center}
\caption{The level diagram of $^{87}$Rb. A naturally occurring gas of $^{87}$Rb would be in a mixture of its 8 electronic ground states, in the 5S manifold. At low magnetic field, these states are well characterized by the total atomic angular momentum, with quantum number $F$, and the magnetic quantum number $m_F$.  Ground-state hyperfine interactions split the $F=1$ and $F=2$ levels by a frequency of $\Delta E_\mathrm{HF}= h \times $6.8 GHz.  Laser cooling, trapping, and imaging typically involve optical transitions to the 5$P$ states.  These are split into two disparate lines, labeled D1 and D2, by the fine structure interaction, and split further into states of distinct total angular momentum $F^\prime$ by the hyperfine interaction.  In $^{87}$Rb, these excited state hyperfine levels are split by about 100 MHz between neighboring lines, greater than the natural linewidth (6 MHz) of the optical transitions.  For some other alkali elements, the excited state hyperfine structure is not resolved.  Especially important are the two ``cycling'' transitions, shown with dashed arrows, on the D2 line, for which spontaneous emission from the excited state cannot return the atom to an electronic ground state other than the one being driven.  Such cycling transitions are favored for laser cooling and optical probing since thousands of photons can be scattered by the atom without its departing from the set of states being interrogated.}
\label{fig:rbspectrum}
\end{figure}

For illustration, the atomic level structure of $^{87}$Rb is shown in Fig.\ \ref{fig:rbspectrum}.  For this alkali atom, strong transitions are found at optical wavelengths of 780 and 795 nm (per Fraunhofer, these are the D2 and D1 lines, respectively).  Each of these lines is split into well-resolved transitions between the various ground- and excited-state hyperfine levels.

\begin{figure}[ht]
\begin{center}
\includegraphics[width=13.5cm]{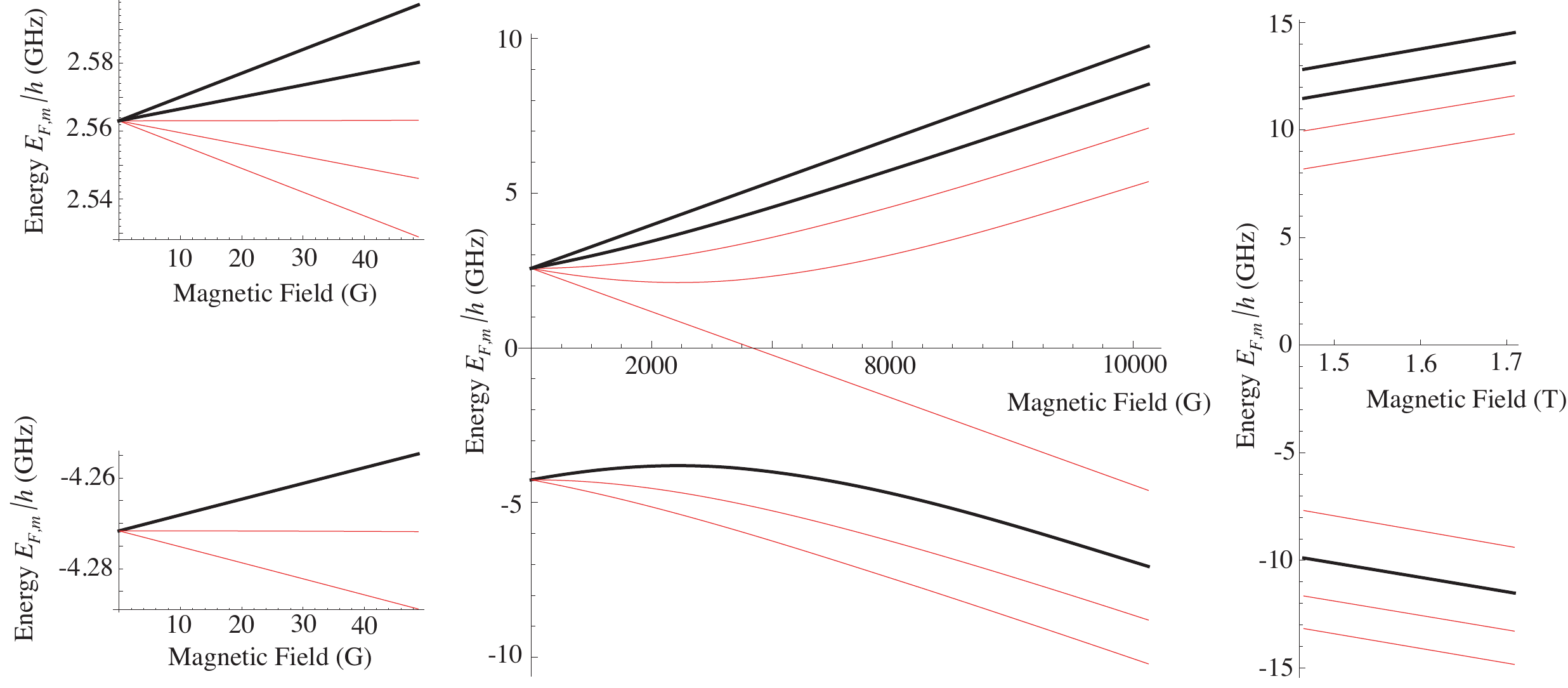}
\end{center}
\caption{Breit-Rabi diagram for $^{87}$Rb. {\bf Left column:} At low field, states are well characterized by their total angular momentum $F$, and their energies deviate linearly from their zero-field hyperfine splitting by $g_F m_F \mu_B B$. States that can be trapped at low field are indicated with a thick black trace. {\bf Right column:} At high field, the energies are again linear in field, varying as $g_J m_J \muB B$. {\bf Central column:} Between these two linear regimes is a quadratic regime in which neither $F$ nor $m_J$ are good quantum numbers.}
\label{fig:RbBR}
\end{figure}
\subsection{Magnetic trapping}
\label{sec:Btraps}

Applying a magnetic field to the atom adds another term to the Hamiltonian, $\hmd = -\hat\mu \cdot \mathbf{B}$, where $\hat\mu$ is the magnetic dipole operator and $\mathbf{B}$ is the external field. This term breaks the rotational symmetry of the Hamiltonian of an isolated atom.  Solving for the eigenvalues of the electronic ground and excited states in an applied magnetic field is a standard exercise in an atomic physics course.  For the case of a $J=1/2$ electronic state, such as the ground-state of alkali atoms, the energy eigenvalues are expressed conveniently by the Breit-Rabi formula,
\begin{equation} \label{eq:BR}
\frac{E_{F, m}}{\Delta E_\mathrm{HF}} =  -\frac{1}{4 F_+} \pm \frac{1}{2} \sqrt{ 1 + 2 \frac{m}{F_+} x + x^2 },
\end{equation}
where $x = g_J \muB B/\Delta E_\mathrm{HF}$, $m=m_I \pm 1/2$,  $\Delta E_\mathrm{HF}=A_\mathrm{HF} F_+$, $A_\mathrm{HF}$ is the hyperfine constant, and $F_+=I+1/2$.
Fig.\ \ref{fig:RbBR} shows the the Breit-Rabi level diagram for the ground state of $^{87}$Rb.  We point out several of its features.  For weak magnetic fields, the magnetic dipole term $\hmd$ can be treated as a perturbation for which the total atomic angular momentum $F$ is still conserved.  The Zeeman energy shifts are then linear: $E \approx E_{B=0} + g_F m_F \mu_B B$, where $m_F$ is the magnetic quantum number $m_F$ (projection of $F$ along the field axis), $\mu_B$ is the Bohr magneton, and the factor $g_F$ is given by Land\'{e}'s formula.

As the field is increased, the field-induced coupling between hyperfine spin states first becomes evident in a quadratic Zeeman shift, proportional to $(m_F \mu_B B)^2 / \Delta E_\mathrm{HF}^2$.  For example, for the $F=1, m_F = \pm 1$ levels $^{87}$Rb, this shift is 70 Hz per square Gauss.  The quadratic Zeeman shift plays an important role in the physics of spinor Bose gases, as discussed in the Chapter by Fetter and Foot.
 	
At very high fields, the magnetic dipole term dominates the energy eigenstructure.  With the hyperfine interaction now treated as a perturbation, the atomic levels become increasingly well described by the quantum numbers $m_J$ and $m_I$, these being the individual spin projections of the electron and nuclear angular momenta.  Since the magnetic moment of the electron is much larger than that of the nucleus, the energy states break up into a higher-energy set of $m_J = +1/2$ eigenstates and a lower-energy set of $m_J=-1/2$ eigenstates, as shown in Fig.~\ref{fig:RbBR}. In this regime,  $E \approx m_J g_J \muB B + A_{\rm HF} m_J m_I$. In each $m_J$ set, level spacing is nearly constant at $A_{\rm HF}$, with a weak dependence on magnetic field.  This slow variation becomes useful in performing rf spectroscopy on resonantly interacting Fermi gases of $^6$Li, for instance, to observe interaction shifts from the single-particle transition energy $A_{\rm HF}/2$.

The variation of atomic energies in an applied magnetic field finds important application in the magnetic trapping of ultracold gases.  To achieve such trapping, atoms are placed in an inhomogeneous magnetic field.  Assuming an itinerant atom follows the magnetic field orientation and magnitude adiabatically, the spatial variation of the magnetic field magnitude leads directly to a spatially varying magnetic potential energy.  Traps are formed typically with fields in the low-field regime discussed above, and, since magneto-static field maxima in conductor-free regimes are not allowed by Maxwell's equations, only ``weak-field-seeking'' atomic states, i.e.\ those with $m_F g_F > 0$, can be trapped magnetically.  For example, for $^{87}$Rb, only the $\ket{F, m_F} = \ket{2,2}$, $\ket{2,1}$ and $\ket{1,-1}$ states are trapped at the limit of zero magnetic field\footnote{In high fields, the quadratic Zeeman effect has been used to trap the $\ket{2,0}$ state of $^{87}$Rb \cite{bouy02CW}.} (see Fig.\ \ref{fig:RbBR}).

The simplest trap is a magnetic quadrupole field, whose minimum field is zero. However, spin-flip losses occur when atom pass too close to the zero-field point. Long-lived magnetically trapped samples use a Ioffe-Pritchard configuration \cite{gott62,prit83}, in which a two-dimensional magnetic quadrupole is combined with a magnetic dipole along the third axis. To lowest order, the combined field is
\begin{equation}
\vc{B}(x,y,z) = B_0 \begin{bmatrix} 0  \\ 0 \\ 1 \end{bmatrix} + B' \begin{bmatrix} x  \\ -y \\ 0 \end{bmatrix} + B'' \begin{bmatrix} - xz  \\ -yz \\ z^2 - x^2/2 - y^2/2 \end{bmatrix},
\end{equation}
where $B_0$ is the holding field, $B'$ is the quadrupole gradient, and $B''$ is the dipole field curvature. For $x,y \ll B_0/B'$, the potential is harmonic in all directions, with frequencies $\omega_z^2 = 2 \muB B'' / M$ and $\omega_{x,y}^2 = \muB (B')^2 / M B_0 - \omega_z^2 /2$. The depth of the trap is limited by off-axis saddle points.

The maximum magnetic dipole moment of an atom is about one Bohr magneton per unpaired electron.  Thus, in a trap where the magnetic field strength varies from zero at the trap bottom to, say, 0.1\,T at the rim of the trap potential, a one-$\mu_B$ atom experiences a trap depth of about 100\,mK, a temperature that is attainable by a precursor stage of laser or perhaps buffer-gas cooling.  Transition-metal atoms that have been recently laser-cooled can have rather large magnetic moments, e.g.\ 6~$\mu_B$ for Cr or 10~$\mu_B$ for Dy, coming from the large number of unpaired electrons in the partially filled shells of these atoms.  However, the large dipole moment of these atoms also makes the weak-field-seeking states more unstable to dipolar relaxation collisions wherein the magnetic potential energy is converted to kinetic energy due to transitions among Zeeman states.  Thus, magnetic trapping of these high-spin atoms has been less useful than for hydrogen-like atoms.  The ground states of alkaline earth atoms, with two paired electrons, lack an electronic magnetic moment, and thus cannot be magnetically trapped.

\subsection{Electrostatic and optical trapping}
\label{sec:Etraps}

Even though neutral atoms with unpaired electrons have a {\it magnetic} dipole moment, eigenstates of $\hamio$ do not have an {\it electric} dipole moment. However, an applied electric field can mix eigenstates of opposite parity to induce a dipole moment.  Barring degeneracies between states of opposite parity (such as occurs for excited states of hydrogen), the resultant Stark shift is quadratic in field strength.  For heteronuclear (``polar'') molecules, the existence of nearly degenerate states of opposite parity means that the Stark shift becomes linear beyond some relatively weak polarizing field.

Restricting our discussion to the quadratic Stark shift typical of ground-state atoms, static electric fields create an attractive potential (the dc Stark shift) with a strength $-\frac{1}{2}\alpha_0 {\cal E}^2$, where $\alpha_0$ is the dc polarizability and ${\cal E}$ is the electric field strength. Alkali atoms have $\alpha_0 \approx 3 \times 10^{-39}$\,Cm$^2$/V. Since fields greater than $10^5$\,V/m typically cause electrode discharge, static potentials cannot be larger than a few $\mu$K, in temperature units. This potential depth is roughly $10^5$ times shallower than the typical magnetic trap depths discussed in Sec.~\ref{sec:Btraps}. Another problem with electrostatic traps is that field maxima are (again) not allowed by Maxwell's equations, so an electrostatic potential must be combined with another type of potential to form a stable trap for atoms.

Fortunately, the Stark potential can be greatly enhanced near optical resonances.  Considering just a ``two-level atom,'' i.e.\ where we can consider only the coupling between one ground state $\ket{g}$ and one excited state $\ket{e}$ of the atom, the expectation value of the dipole operator $\mathcal{H}_{E1} = -\hat{d} \cdot \mathbf{\cal E}$ is
\begin{equation} \label{eq:OmegaR}
\hbar \Omega_R = \la e | \mathcal{H}_{E1} | g \ra = - \la e | \hat{\vc{d}} | g \ra \cdot {\vc{\cal E}},
\end{equation}
where $\Omega_R$ is the Rabi frequency. On resonance, population oscillates between ground and excited states at the frequency $\Omega_R$. Far from resonance, when the detuning $\delta \equiv \omega - \omega_0$ between the driving frequency $\omega$ and the atomic transition frequency $\omega_0$ is much larger than the transition line width $\Gamma$, the ground state feels a second-order energy shift (sometimes called the ``ac Stark shift''), with the value
\begin{equation} \label{eq:shift}
V_g = \hbar | \Omega_R |^2 / 4 \delta.
\end{equation}
As in the static case, this energy scales like ${\cal E}^2$, but now we identify this as the intensity of an electromagnetic wave.  When $\omega \ll \omega_0$, i.e.\ as we approach the limit of the dc Stark shift, the above formula must be corrected by the addition of counter-rotating terms \cite{cohe92}.  Comparing the dc and ac Stark shifts, we see that the resonant enhancement of the Stark shift is approximately $\omega_0/\delta$. In principle, since optical resonances can have a quality factor $\omega_0 / \Gamma$ of $10^8$, a tremendous enhancement is possible. In practice, optical traps are tuned many line widths from resonance to avoid heating from light scattering, so $\omega_0/\delta$ is typically tens to hundreds.

We see that the sign of the ac Stark shift varies with that of the detuning of the drive field.  Referring to the case of $^{87}$Rb and considering only the levels shown in Fig.\ \ref{fig:rbspectrum}, light at a wavelength below 780 nm is blue-detuned with respect to both the principal transitions.  Thus, a ground-state $^{87}$Rb atom exposed to such light experiences a repulsive ac Stark potential, and will be attracted to low-intensity spots of the light field.  In contrast, light at a wavelength above 795 nm is red-detuned with respect to all transitions from the ground-state, so that $^{87}$Rb atoms experience an attractive ac Stark potential and are pulled toward the high-intensity regions.  Light with a wavelength between those of the D1 and D2 transitions can yield either attractive or repulsive potentials depending on the relative strengths of coupling to the two transitions.  These relative strengths depend on the atomic hyperfine and Zeeman state and the polarization of the light field.  Such a dependence allows the ac Stark shift to be used to create state-dependent potentials or to exert fictitious magnetic fields.

The ac Stark potential allows one to trap quantum gases optically \cite{stam98odt}.  The simplest optical trap is formed by a focused Gaussian beam of red-detuned light.   A 10\,W beam focused to a minimum beam waist of 20\,$\mu$m creates an electric field strength of $3 \times 10^6$\,V/m. Such a beam at a wavelength of 1064\,nm induces a trap depth of 2.4\,mK for rubidium. In general, we see that optical potentials can be deeper than electrostatic potentials, but not as deep as typical magnetic traps.

Optical traps have the tremendous benefit of allowing one to trap {\em any} magnetic sub-level of the ground state manifold. For $^{87}$Rb, for instance, there are eight states among the $F=1$ and $F=2$ hyperfine levels. As discussed above, only three of these can be magnetically trapped with a linear Zeeman effect, whereas an optical trap can hold all eight. A further implication is that the magnetic field can be varied arbitrarily if the atoms are trapped optically. This will have important implications in Sec.~\ref{sec:feshbach}.

\subsection{Optical lattices}
\label{sec:lattices}

In a solid, electron band structure arises from the periodic potential created by ions in a
crystalline arrangement. A similar potential landscape can be created for cold atoms by interfering multiple laser beams. Two intersecting plane-wave beams of the same optical frequency create an intensity pattern that varies sinusoidally in space.  Following Eq.\ (\ref{eq:shift}), this intensity pattern produces a periodic potential of the form
\begin{equation} \label{eq:lattice1D}
V_g(x) = V^x_0 \cos^2{(q x)}.
\end{equation}
where $V^x_0$ is the magnitude of the potential, and where $\mathbf{q} = q \hat{x}$, the difference between the wavevectors of the intersecting beams, defines the $\hat{x}$ axis.  A single pair of beams creates a stack of pancake-like potentials, with tight confinement along one axis due to the spatial interference term and loose confinement along the other axes defined by the radial profile of the intersecting laser beams.  When the intersecting laser beams are weak, the atoms may tunnel between neighboring pancakes over experimentally reasonable time scales. When the laser beams are sufficiently intense, such tunneling is negligible, and, at sufficiently low temperature, the atomic motion may be reduced to just two dimensions.

Starting from this basic configuration, a wide range of trapping potentials can be generated. Adding more laser beams, one can create periodic confinement also in the $y$ and $z$ directions.  The simplest three-dimensional lattice potential is one in which three pairs of beams are used along orthogonal axes, with each pair at a different optical frequency so that cross-interference between pairs can be ignored.  The resulting potential is
\begin{equation} \label{eq:lattice3D}
V_g(x) = V^x_0 \cos^2{(k_x x)} + V^y_0 \cos^2{(k_y y)} + V^z_0 \cos^2{(k_z z)}.
\end{equation}
The intensities $V^i_0$ in this cubic lattice can be chosen arbitrarily, and even dynamically, to produce atomic gases that live effectively in three-, two-, one-, or zero-dimensional spaces.  Even the wave vectors can be varied dynamically, in an ``accordion'' fashion \cite{peil03,will08dynamic} or to create a rotating lattice \cite{tung06pin,will08dynamic}. Although square nets and cubic lattices have dominated cold atom experiments to date, all of the Bravais lattices can be created \cite{pets94}.

So far we have tacitly assumed that lattice potentials were created with parallel linear polarizations of light. However the polarization of each laser beam can be chosen arbitrarily. The intensity at $\vc{r}$ of two interfering beams with wave vectors $\vc{k}_{1,2}$ and complex polarizations $\ephat_{1,2}$ is
\begin{equation}
\frac{I_\mathrm{sw}(\vc{r})}{I_0} = 2 + 2 \mbox{Re} \{ \ephat_1 \cdot \ephat_2^* e^{i (\vc{k}_1-\vc{k}_2) \cdot \vc{r}} \}
\end{equation}
where $I_0$ is the intensity of each individual beam. However it is not the intensity alone, but also the optical polarization that determines the potential, as follows from the definition of the Rabi frequency in  Eq.~(\ref{eq:OmegaR}).  Treating the ac Stark effects more completely, accounting for both diagonal and off-diagonal second-order coupling between the different spin states of the electronic ground state, leads to a diverse set of optical lattice potentials, such as spin-dependent lattices, where the external potential depends on the internal state of the atom \cite{deut98}, or lattices with significant spin-orbit coupling, in which bands can have non-zero Chern number \cite{coop11flux}.  Dynamic control of the relative polarization can be used for spin-dependent transport and, thereby, to generate widespread entanglement between the spin and motion states of lattice-trapped atoms \cite{mand03trans,fran06}.  Considering also the possibility of optical superlattices (formed by overlaying lattices of different periodicity), lattices for atoms with several stable electronic states (notably the alkali-earth atoms), lattices that combine both optical and radio-frequency manipulation of the trapped atoms, and so forth, one might expect that atomic physicists will be as busy studying lattice trapped atoms as condensed-matter physicists have been studying crystal bound electrons!

However, as one considers designing a new family of optical lattice potentials, one should keep in mind the limitation that optical potentials are not entirely conservative.  The electric dipole moment induced dynamically in an atom by a near-resonant laser field does not respond exactly in phase with the laser drive.  The out of phase (imaginary) electric susceptibility of the atom induces the atom to emit light spontaneously.  The recoil from these randomly emitted photons and also the random transitions between electronic states of differing polarizability (dipole-force fluctuations) heat the atom mechanically \cite{gord80}.  Many of the machinations that yield the most intriguing optical lattice potentials often require the use of laser light that is fairly close to atomic resonances, so that the laser detuning from various transitions be significantly different.  Perhaps realizing the more exotic optical lattice configurations will have to wait for more exotic atoms and molecules to be introduced to the ultracold regime.

%
%

\section{Interactions}
\label{sec:interactions}

As discussed above, the broad relevance of research on quantum gases rests upon the idea that the many-body quantum systems behave similarly regardless of the details of their construction.  Yet, in considering analogies between electrons in solid-state materials and atoms in artificial potentials, one cannot escape the fact that electrons, being charged, interact via the long-range Coulomb interaction, while neutral atoms do not.  Rather, in most neutral atomic gases, interactions are short-ranged as defined by the Angstrom-scale effective range of the two-atom molecular potential. The absence of long-range interactions among atoms does not necessarily invalidate the analogy between atomic and electronic materials, since in many materials, the Coulomb interactions among mobile electrons are screened at long range, leaving only an effective short-range interaction between electron-like (or hole-like) quasiparticles.  However in other contexts, the absence of electric charge and long-range interactions lead to a major departure of ultracold atom systems from typical solids, for instance \emph{vis a vis} the connection between plasma oscillations and Bogoliubov modes in charged superconductors or the realization of high-order and long-range spin-spin interactions in magnetic materials. The presence of dipolar interactions in some neutral gases is interesting exception to the rule, and is discussed at the end of this section.

\begin{figure}[tb!]
\begin{center}
\includegraphics[width=7cm]{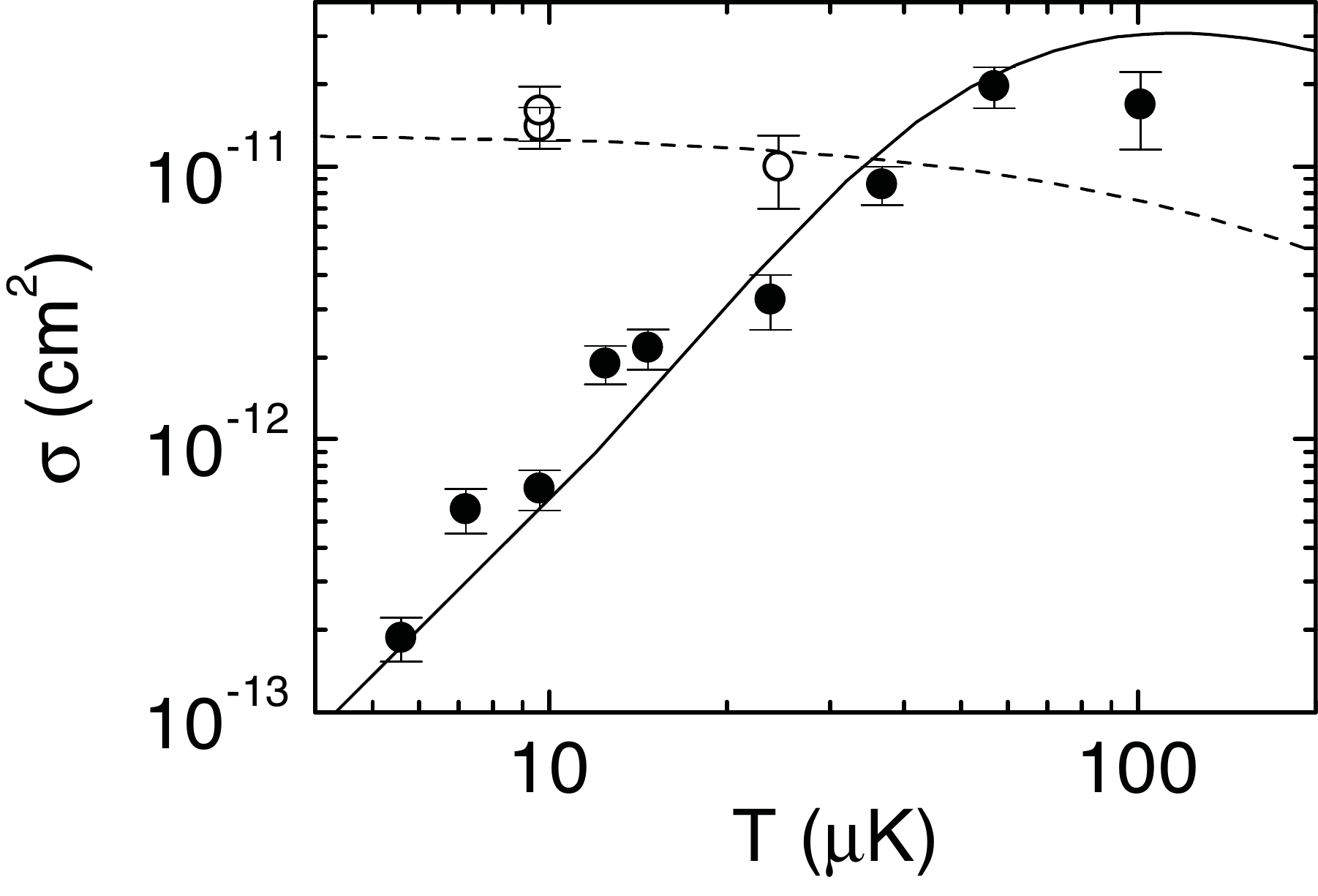}
\end{center}
\caption{The s-wave collision cross section (open symbols), measured using a mixture of spin states of fermionic $^{40}$K, shows little temperature dependence. However the p-wave cross-section (closed symbols), measured using spin-polarized atoms, exhibits a threshold behavior, falling two orders of magnitude between 100\,$\mu$K and 5\,$\mu$K. Reproduced from Ref.~\cite{dema99pwave} with permission, courtesy of the APS. }
\label{fig:FreezeP}
\end{figure}

\subsection{The scattering length}
\label{sec:as}

At the low density of atomic gases, interactions can be treated as a two-atom quantum scattering problem governed by the relevant molecular potential.  A complete treatment of such scattering is quite complicated, owing to the many potential-energy surfaces that arise as two compound quantum objects -- atoms with internal states describing electronic motion, electron spin, and nuclear spin -- approach one another.  However, much of this complexity is swept under the rug due to the near-zero incident energy of a colliding atom pair.  One appeals to the partial wave treatment of scattering, in which the incident matter waves engaged in a collision are decomposed into eigenstates of angular momentum.  In this basis,  a centrifugal barrier exists for any non-zero angular momentum, for example p-wave or d-wave collisions. However as shown in Fig.\ \ref{fig:FreezeP}, even for p-wave collisions, the magnitude of the barrier is $\sim$0.1\,mK, much higher than the sub-$\mu$K temperatures of quantum degenerate gases. Therefore only isotropic s-wave scattering is significant for ultracold atoms.  Far from the innards of the molecular potential, such s-wave scattering results only in a phase difference, $\eta_0$, between the incoming and outgoing deBroglie waves, which is given as \cite{chin10rmp}
\begin{equation}
k \cot{(\eta_0)} = - \frac{1}{a_s} + \frac{1}{2} r_{e} k^2,
\end{equation}
where the wavevector $k$ relates to the relative momentum of the colliding pair.  In the ultracold regime, all the gory details of the interaction potential may be neglected (though sometimes it pays to keep them in mind) by retaining just two quantities that characterize the scattering phase: the s-wave scattering length $a_s$, and the effective range of the potential $r_e$. The effective range term itself is negligible for collision energies much less than $\hbar^2 / M a_s r_e$, which is roughly 100\,$\mu$K for $^{87}$Rb. For $^{87}$Rb in the polarized $\ket{2,2}$ state, the triplet scattering length $a_s=5.238(3)$\,nm \cite{vankemp02}.

The scattering length also determines the elastic collision cross section, given for distinguishable particles as
\begin{equation}
\sigma_\mathrm{NI} = \frac{4 \pi a_s^2}{1 + k^2 a_s^2}.
\end{equation}
For identical bosons, there is an additional factor of two: $\sigma=2 \sigma_\mathrm{NI}$, whereas for identical fermions, s-wave collisions are forbidden by the Pauli principle, and $\sigma=0$. Figure~\ref{fig:FreezeP} shows direct measurement of the latter effect in a spin-polarized gas of $^{40}$K.

Since the details of the interaction potential are unimportant, it is convenient to replace it formally with a simpler potential -- the zero-range contact potential $V(\vc{R})=g \delta(\vc{R})$ with $\vc{R}$ being the relative spatial coordinate -- which yields the same s-wave scattering length by identifying $g=4 \pi \hbar^2 a_s / M$.  Subtle divergence problems that arise sometimes from this pseudo-potential approach are avoided by using regularized potential, $V(\vc{R}) \Psi(\vc{R})=g \delta(\vc{R}) \partial_R(R \Psi(\vc{R}))$ \cite{huan57}.

From dimensional arguments, and in mean-field theory, it follows that the product $n g$, with $n$ being the number density, quantifies the per-atom interaction energy.  This energy is seen to be proportional to the scattering length, i.e.\ we associate $a_s>0$ with repulsive interactions and $a_s<0$ with attractive interactions.  The possibility of realizing both attractive and repulsive interactions may seem counter-intuitive, given that the interaction energy between polarizable ground-state atoms is always \emph{negative} at long-range (at the typical distance between atoms) due to van der Waals interactions.  More properly, the per-atom interaction energy should be thought of as the influence of scattering on the \emph{kinetic energy} of the interacting gas.  To visualize this, consider two atoms trapped in a box with linear dimensions $l$, which we assume to be large.  Their kinetic energy goes as $\hbar^2 / M l^2$.  Due to scattering, the two-body wavefunction behaves at long range as if the linear dimension of this box now has a length $l + a_s$.  For $a_s>0$, the volume inhabited by a collection of atoms is effectively reduced due to the asymptotic effects of scattering, causing their kinetic energy to rise.  For $a_s<0$, adding more atoms seemingly \emph{increases} the volume available to the gas, causing its kinetic energy to diminish.

\begin{figure}[tb!]
\begin{center}
\includegraphics[width=8cm]{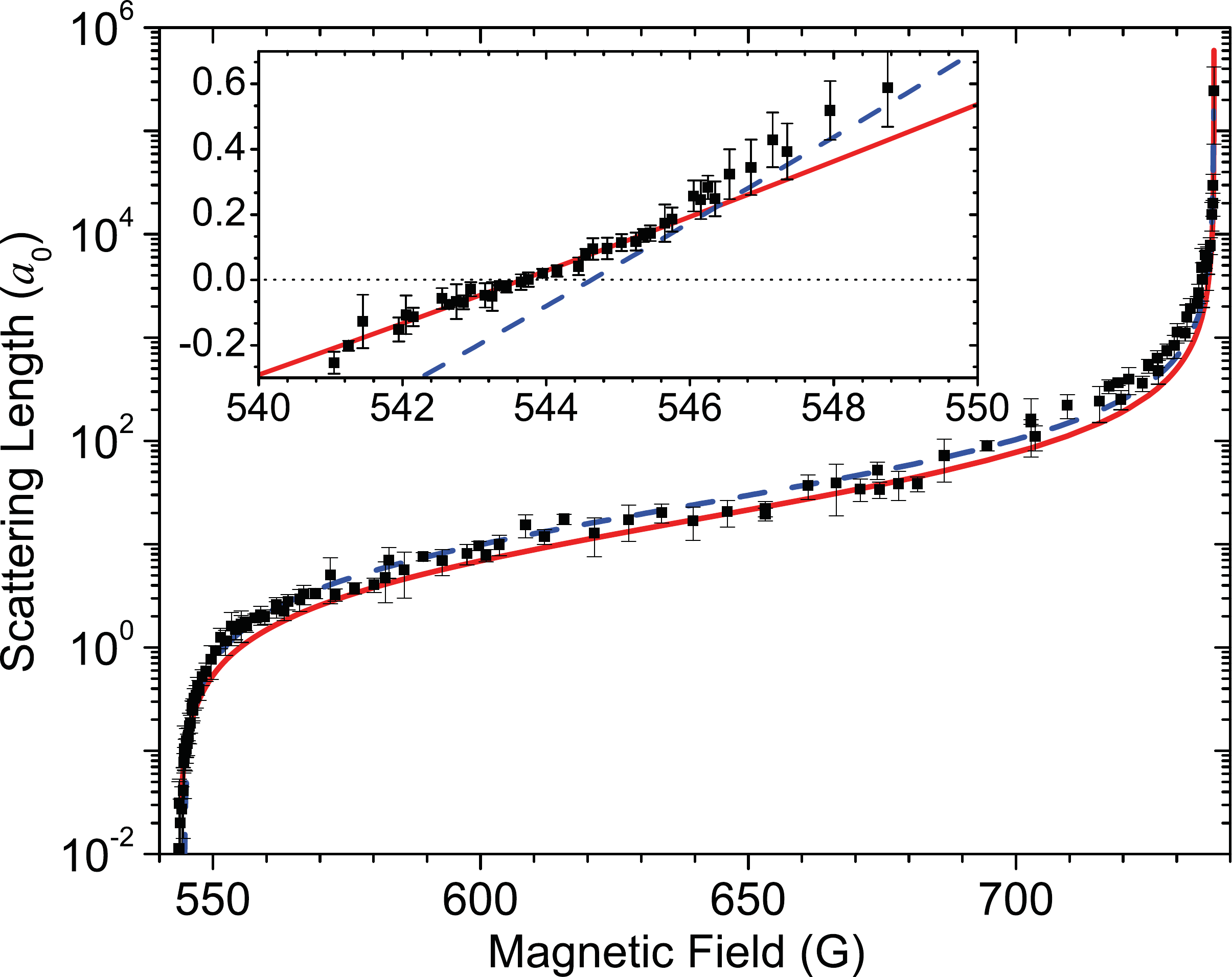}
\end{center}
\caption{A Feshbach resonance at 737\,G is used to tune the scattering length of bosonic $^7$Li across seven decades, from extremely small attractive interactions to extremely strong repulsive interactions. In-situ cloud size is used to measure the scattering length (points). The mean and standard error of approximately ten shots taken at each field is shown. Results of a coupled-channels calculation are shown by the solid line. The Feshbach resonance fit is indicated by the dashed line. The inset shows the extracted values of $a_s$ near the zero crossing.  Values of the scattering length above $10^4$ Bohr radii are not accurate due to beyond mean-field effects.  Reproduced from Ref. \cite{poll09extreme} with permission, courtesy of the APS. }
\label{fig:tuning}
\end{figure}

The treatment of three-dimensional scattering discussed above is modified in fascinating ways by the confinement of atoms to lower-dimensional spaces, through which interaction strengths are rescaled and various confinement-induced collisional resonances and bound states can occur.  These effects are discussed in the Chapter by Hadzibabic and K\"{o}hl.

\subsection{Feshbach resonances}
\label{sec:feshbach}

In the presence of a magnetic field, free atoms and bound molecules experience a differential Zeeman shift since their magnetic dipole moments are not equal. At certain serendipitous values of the field, this differential energy shift can bring a bound dimer state into resonance with the energy of two asymptotically free atoms. This condition is called a Feshbach resonance \cite{chin10rmp}. Even though the atoms do not actually form a dimer, the phase shift of their s-wave collision is modified by the proximity of the resonance. As shown in Fig.\ \ref{fig:tuning}, by using a magnetic field, the ultracold gas can be tuned from weakly to strongly interacting, and chosen to be either repulsive or attractive.

Such an exquisite control of interactions is possible with cold atoms because the interactions are binary. Liquid helium is roughly a billion times more dense, meaning that particles are a thousand times closer to one another. At the microscopic level, it's an ``atomic traffic jam'' --- the mean separation is comparable to the range of the interatomic potential, and no clean separation of energy scales is possible. For neutral gases, the Feshbach resonances can be milli-Gauss wide, showing that the control over interactions through delicate tuning of the two-body molecular potential is not compromised by many-body effects.

Experiments with Feshbach resonances use optical traps, for two reasons. First, Feshbach resonances typically occur at fields of hundreds of Gauss ($\sim 10$\,mT), whereas the magnetic-field minima of magnetic traps are often at the single-Gauss level ($\sim 0.1$\,mT). Second, the Feshbach resonances are accessed at the lowest magnetic fields in anti-trapped Zeeman states. As discussed in Sec.~\ref{sec:Btraps}, $m_F g_F < 0$ atoms cannot be trapped magnetically.

An alternative to the magnetic Feshbach resonance is the coupling to molecular states with electric fields, either static \cite{mari98,krem06controlling} or optical \cite{fedi96,thei04feshbach}. These techniques are especially important for atoms with paired electrons, such as rare earth alkalis, since the hyperfine interaction couples free states to bound states in magnetic Feshbach resonances.  Optically induced Feshbach resonances are being explored as a way to produce novel systems with rapid temporal and spatial variations in the interaction strength.  To date, however, these optical Feshbach resonances appear to be accompanied by the significant disturbance of the gas by spontaneous light scattering.

Sweeping the magnetic field across a Feshbach resonance can cause the adiabatic binding of free atoms into the loosely bound molecular states which induce the resonance.  These ``Feshbach molecules'' are quite unusual in that they are bound by mere kHz-level binding energies ($10^{-7}$ wavenumbers!) and have sizes in the nanometer range \cite{kohl06rmp}.  The generic few-body physics of Feshbach-resonating atoms has been explored in a number of landmark experiments with cold atoms.  Feshbach molecules are also centrally relevant to studies of the BCS-BEC crossover in degenerate Fermi gases, across which Bose-condensed Feshbach molecules transform smoothly into Cooper pairs in a BCS superfluid.  The formation of Feshbach molecules is also an important stepping stone for the creation of ground-state molecules by photo-association.

\subsection{Dipolar interactions}

The magnetic dipole moment we discussed above for magnetic trapping can also lead to inter-atomic interactions. Atoms in excited states (especially in highly excited Rydberg states) can have strong electric dipole moments, but they are typically not stable when colliding. Molecules offer the possibility of strong electric dipole moments for both trapping and interactions. 

Uniformly oriented dipoles interact according to the potential
\begin{equation}
V_{\mathrm{dip}}(\mathbf{r}) = \frac{C_{\mathrm{dd}}}{4 \pi} \frac{1 - 3 \cos^2\theta}{r^3}
\end{equation}
where $\theta$ is the angle between the dipole orientation and $r$ the separation between particles.  A potential scaling as $1/r^\xi$ at large distance is defined as short-ranged --- and can be treated using the pseudo-potential approach (Sec.\ \ref{sec:as}) --- so long as $\xi>D$, where $D$ is the number of dimensions. For instance, van der Waals interactions arising from fluctuation-induced dipole moments fall off like $1/r^6$ (or faster). Thus, such interactions can be summarized by the scattering length $a_s$ and otherwise ignored. However the $V \sim 1/r^3$ interaction between permanent dipole moments is, by this definition, long-ranged in three dimensions, and must be treated by a different approach \cite{laha09review}.  For neutral atoms, ``dipolar interactions'' is nearly synonymous with ``long range interactions.''

One can compare the strength of the magnetic dipole interaction to that of the s-wave contact interaction by calculating the magnetic self-energy of a trapped gas to the contact interaction strength $g$ (Sec.\ \ref{sec:as}), yielding the ratio $\epsilon_\mathrm{dd} = C_\mathrm{dd} / 3 g$.  For the alkali gases, with a typical s-wave scattering length of tens of Bohr radii and a magnetic moment of one $\mu_B$, $\epsilon_\mathrm{dd}$ is typically below $10^{-2}$, so that magnetic dipole interactions are small compared to the overall contact interaction energy of a single-component Bose or Fermi gas.  However, dipolar interactions may still be significant in determining spin ordering in multi-component alkali gases.  Most notably, for $^{87}$Rb, the contact interaction energy varies only slightly for different magnetic states, as determined by a small difference in s-wave scattering lengths for collisions among different pairs of atomic spin states.  In face of this small variation, the magnetic dipolar energy may indeed be significant \cite{veng08helix}.

More dramatic effects of the magnetic dipole interaction are seen in atomic gases with larger magnetic moments, and in which the s-wave scattering length can be tuned near zero.  Seminal experiments on such ``quantum ferrofluids'' were performed using Cr gases held in anisotropic optical traps. Fixing the orientation of the atomic magnetic moment (of magnitude $6 \mu_B$) to that of the guiding magnetic field, and varying the orientation of that field with respect to the axes of the optical trap, the interactions were converted from being effectively repulsive (for atoms spread out preferentially in directions transverse to the dipolar axis), allowing large condensates to remain stable, to attractive, whereupon the condensates collapsed \cite{laha07ferrofluid}.

Even stronger dipolar effects are expected in other atomic and molecular gases. Neutral atoms in Rydberg states have strong electric polarizabilities due to large-radius electron orbitals, in addition to long-range van der Waals interactions. However the collisional instability of Rydberg atoms prevents a simple comparison to ground-state dipoles.  Heteronuclear molecules polarize in a relatively weak electric field to have an electric dipole moment on the order of 1 debye. When normalized with respect to a typical $a_s$, a dimensionless strength $\epsilon_\mathrm{dd} \gtrsim 10$ is expected.  Quantum degenerate gases of polar molecules have yet to be produced, but several teams are close \cite{ni08highpsd,deig08} and expected to succeed in the near future.

%
%
%
%
%

\section{Taking data}

The constellation of measurement tools available for studies of materials in the solid state reflects the robustness of such materials.  The copious scattering of high-energy photons, electrons and neutrons, injections of decaying muons, and the scratching of solid probes across the surface can all be tolerated by solid materials without necessarily destroying them.  In contrast, cold-atomic materials are delicate.  The momentum transferred to an atom upon scattering just a single near-infrared photon will typically exceed the typical thermal, Fermi, and interaction energy in a quantum vapor. Material probes poking into a nano Kelvin temperature gas will quickly boil the atoms away, ruling out most scanning probe techniques. The delicacy of atomic many-body systems implies that the act of extracting data from such systems is extremely influential on their evolution.  Efficiency is essential, and accounting for the back action of measurements is necessary for their interpretation.

The lion's share of measurements on cold gases are performed optically. The lightly interacting atoms in a quantum gas are strong, narrow-band optical emitters, as discussed in Sec.~\ref{sec:structure}. However ion-based detection can be used for fast detection of single atoms (see Sec.~\ref{sec:singleAtom}), especially for metastable gases that ionize upon impact with an electrode surface.

\subsection{The experimental cycle: the birth and death of an ultracold gas}
\label{sec:cycle}

In conventional condensed-matter physics research, one makes or identifies a material sample, prepares it for measurement, inserts it into a measurement apparatus, and takes measurements for as long one needs to gather reliable data on the phenomenon being investigated.  The same sample might be used for months of data taking or for several sequential types of measurement, and might be stored in a drawer for further investigations down the line.

Ultracold atomic materials are quite different. The material must be constructed prior to each measurement, starting from a new, hot atomic vapor.  The typical experimental sequence is (1) using laser cooling and trapping to gather atoms from the vapor into a magneto-optical trap, with temperatures on the order of 100 $\mu$K, (2) trapping the atoms in a conservative potential, e.g.\ a magnetic (Sec.\ \ref{sec:Btraps}) or optical-dipole (Sec.\ \ref{sec:Etraps}) trap, (3) evaporatively cooling the atoms by gradually lowering the depth of the conservative trap and letting atypically high energy atoms escape the trap, reaching temperatures in the quantum-degenerate regime (usually sub-$\mu$K) and (4) putting the final touches on the material by turning on the system Hamiltonian (interaction strength, lattice type, spin admixture, etc.) that one wants to examine.  Once finally produced, the lifetime of the ultracold material is short, and so the entire experimental sequence for probing the material is quick, typically much shorter than the time it took to prepare the material in the first place.  At the end of each measurement, the sample is discarded.  This cycle is repeated at the cycle time of a few seconds to a few minutes, depending on the speed of the accumulation and cooling stages. In examining a graph of data from a cold atom experiment (For example, Figs.~\ref{fig:FreezeP} or \ref{fig:tuning}), one should value the fact that each point on the graph represents one or several repetitions of a make-probe-discard experimental run.  In light of this protocol, it is somewhat stunning to hear of cold-atom experiments that require hundreds or thousands of ``shots,'' each reproducing a gas under almost identical conditions, to obtain the high precision required to reveal new phenomena or test recent theories.

\subsection{Imaging}

Cold atoms are conventionally probed by optical imaging.  Probe light at a well defined optical frequency is sent through the atomic gas and imaged onto a camera.  Assuming the gas is sufficiently thin that we can neglect double refraction, and neglecting scattering at large numerical aperture, such a probe measures the column-integrated complex dielectric susceptibility tensor, truncated to the two polarizations allowed for light propagating along the imaging axis.  In other words, information about the atomic gas is encoded onto the absorption, phase shift, and polarization state of the optical probe field, properties that can be extracted by appropriate imaging methods.

For example, in \emph{absorption imaging}, one images the probe light directly onto a camera.  Comparing images taken with and without the atomic gas within the field of view, one records the fractional transmission of probe light, $T(\rho)$, for each camera pixel (labeled by position $\rho$).  In the simple case of linear scattering from atoms in a single initial internal state, the areal density of the gas is given by $n_A(\rho) = -\ln T(\rho) / \sigma$, with $\sigma$ being the light absorption cross section given the atomic initial state and the optical polarization and detuning.  Yet, this simple interpretation of the absorption signal may be complicated by various factors such as saturation, optical pumping, polarization rotation, the displacement and acceleration of atoms due to light forces from the probe, collective scattering, inhomogeneous broadening, and others.  Some of these complications are obviated by probing the gas with probe light well above the saturation intensity, so that atoms scatter light at a known maximum rate (one-half the excited-state decay rate), and counting the number of absorbed photons \cite{rein07sat} (or detecting the metered fluorescence \cite{depu00sat}).

An advantage of absorption imaging is that an atom can scatter very many photons in a single imaging pulse, improving the signal-to-noise ratio of the optical measurement.  Indeed, several groups have quieted absorption imaging to the point of measuring atomic distributions with sub-Poissonian noise -- where the atom number $N$ in a given region is measured with uncertainty $\delta N < N^{1/2}$ -- to reveal correlations generated by interatomic interactions or quantum statistics \cite{este08squeeze,sann10suppress,mull10local}.

Absorption imaging has two primary drawbacks.  First, absorption imaging is typically only a single-shot probe.  The photons ``absorbed'' by the atoms in such imaging are, in actuality, re-scattered out of the imaging system.  Such photon scattering imparts significant momentum to atoms within the gas, typically adding sufficient energy and entropy that the atomic quantum system is irrevocably disturbed and must be discarded and created anew for further measurements.  The destructiveness of the measurement makes it challenging to use absorption imaging to characterize interesting temporal dynamics, such as equilibration dynamics, responses to sudden variations in the system Hamiltonian, collective excitations, or temporal noise correlations. If the response of the system to a temporal perturbation is reproducible, one can construct a time-series of measurements from many repeated experiments with variable delay before the absorption probe.  We note it is possible to get around the single-shot limitation by applying absorption imaging to just a small fraction of atoms extracted from the quantum gas. For example, Frelich \emph{et al.} used a weak microwave pulses to transfer $^{87}$Rb atoms to a different hyperfine ground state before imaging them with light that was sufficiently far detuned from the optical transitions of the remaining gas that the trapped sample continued to evolve with little disturbance \cite{frei10vortex}.

Second, absorption imaging is also of limited use in probing gases at high optical densities, where imaging noise leads to large uncertainty in the measured column density.  For this reason absorption imaging is rarely used to measure trapped atoms directly, \emph{in situ}. Rather, absorption imaging is applied typically to gases \emph{ex situ}, after they've been released from their trap and allowed to expand significantly before probing -- a method known as \emph{time-of-flight imaging}.

Alternately, one may use \emph{dispersive imaging} to probe high-density, trapped atomic gases \emph{in situ}.  To attain the highest imaging resolution, one uses probe light sufficiently detuned so that diffraction, rather than refraction, is dominant, a condition achieved when the phase shift imparted on probe light on the order of 1 radian or less.  For high optical density gases, i.e.\ those for which absorption imaging is problematic, this condition also implies that optical dispersion dominates over optical absorption.  Information on the atomic gas is now encoded in the phase of the transmitted light.  This phase is detected by one of several common imaging techniques, such dark-field \cite{andr96}, phase-contrast \cite{andr97prop}, or polarization-contrast \cite{brad97bec} imaging.

\begin{figure}[t]
\begin{center}
\includegraphics[width=3in]{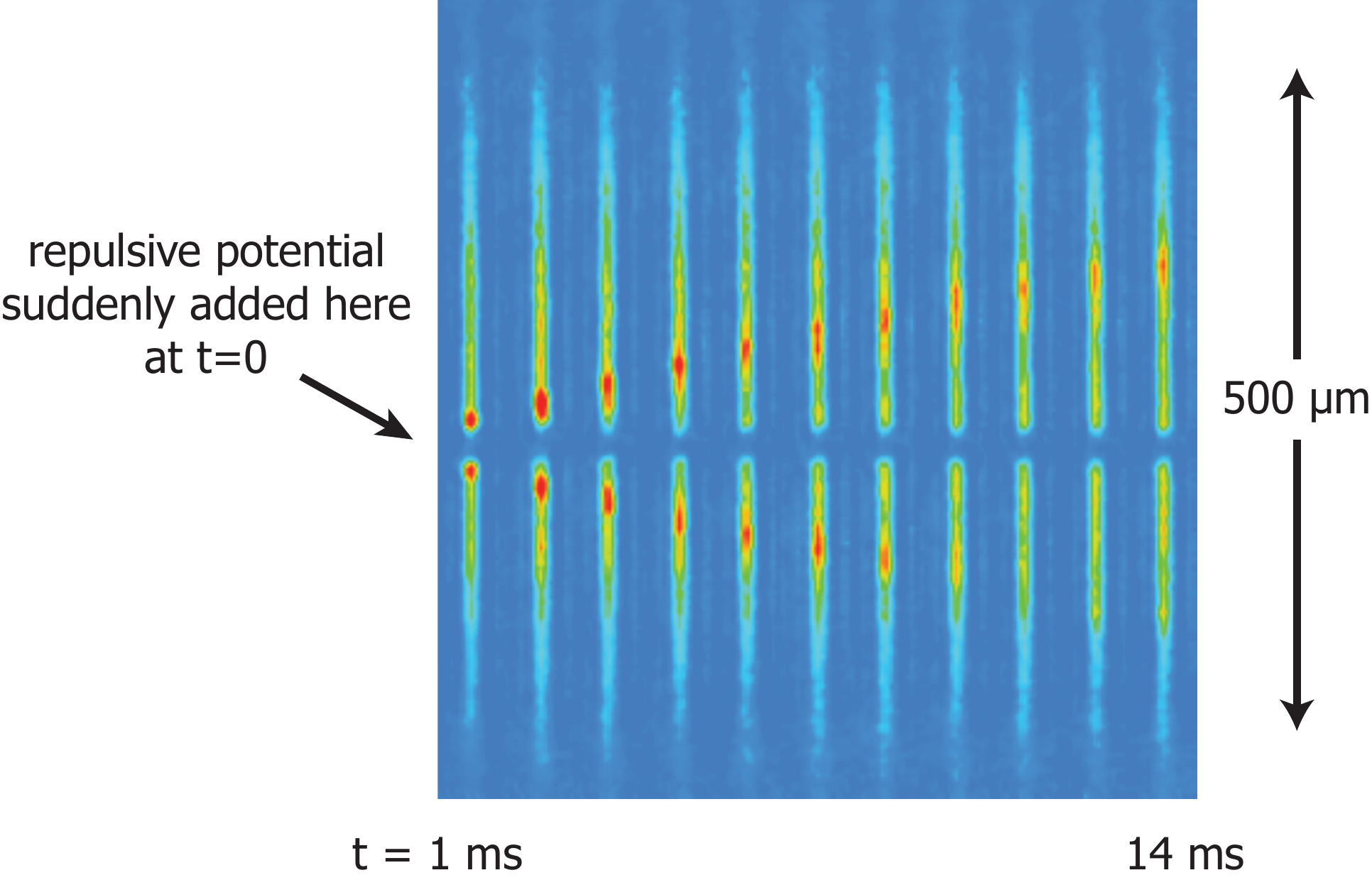}
\end{center}
\caption{Repeated phase-contrast dispersive imaging is used to trace the propagation of a sound wave within a single trapped Bose-Einstein condensate.  At $t=0$, a repulsive potential is introduced suddenly within the middle of the elongated condensate.  Images are taken at 1.3 ms intervals, beginning at $t = 1$ ms.  The density excesses (red) travel along the condensate, revealing the speed of sound.  Figure adapted from Ref.\ \cite{andr97prop}.}
\label{fig:soundprop}
\end{figure}

\begin{figure}[t]
\begin{center}
\includegraphics[width=5 in]{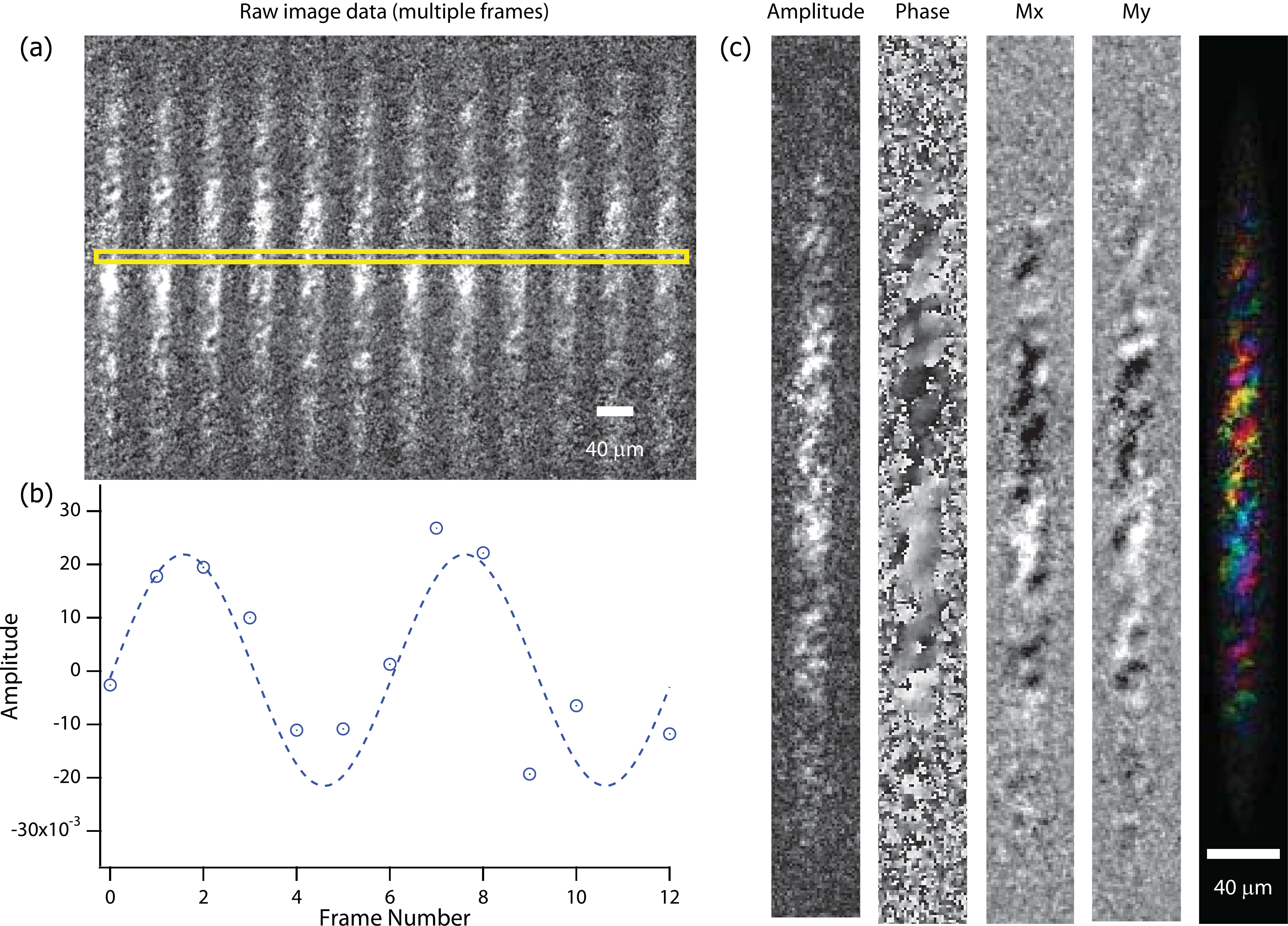}
\end{center}
\caption{Repeated polarization-contrast imaging is used to measure both components of the transverse magnetization of a spin-1 $^{87}$Rb gas. (a) Several brief polarization-rotation images are taken, measuring one component of the magnetization transverse to an applied magnetic field, while such magnetization undergoes Larmor precession between image frames.  Such Larmor precession is evident by selecting data from a common location in each image frame (yellow band in a) and observing the temporal oscillation (shown in b).  (c) Analyzing data from each image pixel, one extracts the spatially varying Larmor precession amplitude and phase, or, equivalently, the two components of the transverse magnetization (here denoted as $M_x$ and $M_y$, represented either in gray scale, or in a color scale with brightness indicating amplitude and hue indicating orientation).  This image analyzes a degenerate gas equilibrating to a ferromagnetic state, as reported in Ref.\ \cite{guzm11}.}
\label{fig:larmor}
\end{figure}

Dispersive imaging is still somewhat destructive, owing to the residual off-resonant absorption by the gas.  The destructiveness is mitigated only by imaging at lower signal-to-noise ratios (e.g.\ by using fewer photons or probe light detuned further from resonance); only so much information can be garnered from the gas before it is destroyed.  Nevertheless, by parsing this information it is possible to obtain several \emph{in situ} images of a gas, e.g.\ to measure the time evolution (Fig.\ \ref{fig:soundprop}) or to measure several properties (Fig.\ \ref{fig:larmor}) of a single sample.

\subsection{Bragg scattering}

The imaging probes discussed above are analogous to transmission electron or x-ray microscopy in which real-space images are obtained of solid-state materials.  Alternately, solid-state materials can be characterized in reciprocal or momentum space, e.g.\ by using Laue or Bragg x-ray diffraction to identify lattice planes and crystal orientations or neutron diffraction to identify magnetic order.  Such scattering experiments on solids are most powerful in the form of angle resolved spectroscopy, by which one determines both the momentum and the energy absorbed by a material sample by scattering a probe particle.

A similar family of methods has been developed to characterize the structure of atomic gases by angle-resolved light scattering.  As for the methods used for solid-state materials, Bragg scattering of light can be used to determine the dynamic and static structure factors of a quantum gas, revealing, for example, the role of interactions in suppressing density fluctuations \cite{stam99phon} or shifting excitation energies \cite{sten99brag,stei02exc,thei04feshbach,papp08}.

A notable difference in the quantum-gas applications is that coherent laser light interacts strongly with the gas and also possesses a wavelength that is on par with the physically relevant length scales of the gaseous sample.  For this reason, experimental applications of \emph{Bragg spectroscopy} typically make use of \emph{stimulated} Bragg scattering to characterize the sample.  Rather than quantifying the momentum and energy transfer by spectroscopic analysis of (few) scattered particles, one specifies such quantities \emph{a priori} and with excellent resolution by the wavevector and energy difference between several plane-wave light beams impinging on the sample.  The excitation strength is then quantified, e.g.\ by detecting the amplification (or suppression) of one of the driving light fields \cite{pino11counting}, or the energy or momentum deposited in the gas \cite{sten99brag}.

The similar method of \emph{modulation spectroscopy} is well suited to the study of lattice-trapped quantum gases.  Instead of the optical plane waves used for Bragg scattering, here, the optical excitation is effected by standing-wave modes of light, these being the optical lattice beams themselves.  The carrier and modulation sidebands of the amplitude-modulated lattice beams define the energy transferred to the medium in transitions at constant quasi-momentum.  Such modulation spectroscopy has been used to characterize the Mott insulator formed by lattice-trapped Bose gases \cite{scho04} and the particle-hole excitations in Fermionic samples \cite{hein11fermi}.

A distinct feature of research on atomic gases is that measurements need not be confined to the perturbative regime.  For example, modulation spectroscopy is performed often by modulating the potential energy (the periodic optical lattice) by tens of percent.  On the one hand, the use of such a non-perturbative measurement complicates its interpretation.  On the other hand, such strong perturbations produce interesting non-equilibrium systems worthy of study in their own right, for example, Bose-Einstein condensates with macroscopic occupancy of a phonon state \cite{katz05} and dynamic systems in which the initial excitation of the system is fed back via matter-wave mixing or optical cavities \cite{inou98,slam07super,baum10dicke}.

\subsection{Single-atom probes}
\label{sec:singleAtom}

The limits of measurement sensitivity are reached when a high-fidelity record of each and every atom of a many-body system is obtained.  Remarkably, recent experimental developments are converging on this limit.  Several approaches are being pursued.  For instance, it is possible to detect each atom in a two- \cite{bakr09microscope,sher10singleatom} or three-dimensional \cite{nels07} array.  To achieve such a measurement, atoms in an optical lattice are exposed to near-resonant light that is scattered copiously by the atoms as they are being actively laser cooled.  Such fluoresced light is then imaged with sufficient spatial resolution to isolate the emission of each atom from that of its neighbor.  High measurement sensitivity has also been attained for metastable atomic species, such as the lowest state of orthohelium \cite{robe01he}.  Upon striking a detector surface, a single metastable atom may undergo Auger decay, releasing electrons that can be fed into a multi-channel electron multiplier.  Other single-atom-sensitive techniques include cavity-enhanced optical detection \cite{ottl05} and multiphoton ionization \cite{kraf07,henk10}.  Ultimately, such methods may allow for a detailed measurement of the state of an interacting many-body quantum system, providing a wealth of information beyond the standard correlation functions that are measured typically for solid-state systems.

\section{Interpreting Data}

As discussed above, the atomic physics toolbox provides ready means to extract data from gaseous systems.  However, to turn such data into a meaningful characterization of system properties, particularly in forms that are comparable to those that characterize solid-state materials, requires another round of ingenuity.  Many of the major recent advances in cold atomic physics represent such ingenuity.  Here we provide a few examples of how the raw data of measurements are interpreted as revealing the underlying physics of atomic quantum gases.

\subsection{Extracting thermodynamic quantities from spatial density profiles}
\label{sec:LDA}

We are often asked, ``how do you measure such low temperatures?''  One measures the temperature, or other thermodynamic property, of a conventional material by connecting it with a reliable and calibrated sensor, which is a second material the properties of which are already fully understood.  While different in practice, the concept behind measuring bulk properties of quantum gases is the same: temperature and pressure are measured by reference to the properties of an ideal gas.

For example, at thermal equilibrium, the velocity distribution of a non-degenerate ideal gas is given by the Boltzmann distribution, $f(\mathbf{v}) \propto \exp\left(- M v^2 / 2 k_B T\right)$.  One may then determine the temperature by measuring how fast the particles in such a gas (of mass $M$) are moving.  In principle, this can be done by seeing how quickly the gas escapes from its container once the trapping potential is turned off.  This simple concept can also be extended to some of the degenerate and interacting systems that are under study today.  One notes that the most highly excited atoms in the gas are less affected by interactions, and so a reliable temperature measurement can still be made by examining the high-velocity tails.

However, with increasing interactions, one cannot rely on the time-of-flight distribution being a true reflection of the \emph{in-situ} velocity distribution: the gas particles scatter off each other during the expansion from the trap, resulting in thermal cooling and a modification of the velocity distribution.  Other approaches to measuring velocity distributions, such as atom interferometry \cite{sten99brag}, also fail in strongly interacting gases as the Doppler and interaction-induced effects are difficult to disentangle.

An alternate approach is to base one's measurements on the \emph{in situ} spatial distribution of the gas.  Analyses of such distributions rely typically on the local density approximation, i.e.\ that even in the inhomogeneity introduced by the non-constant trapping potential across the volume of the trapped gas, properties of the gas measured locally -- namely the density, compressibility, pressure, magnetization, etc.\ -- are interrelated by the equation of state of a homogeneous quantum gas.  In fact, the inhomogeneous trapping potential of the gas becomes the vehicle by which such properties are measured in the first place. The constant chemical potential of an equilibrated gas may be written as $\mu_0 = \mu(\mathbf{r}) + V(\mathbf{r})$ where, by the local density approximation, we interpret $\mu(\mathbf{r})$ as the ``local'' chemical potential that determines local properties of the gas according to its equation of state.  Consider that one measures the \emph{in situ} density profile $n(\mathbf{r})$ over a region where the trapping potential has a non-zero gradient $\nabla V(\mathbf{r})$.  One thereby obtains two thermodynamic quantities simultaneously: $n(\mu)$ and $dn/d\mu = - |\nabla n| / |\nabla V|$.  Together with a means of determining $\mu_0$ and the temperature $T$, one obtains an experimental characterization of the equation of state.

\subsection{What one learns about quantum gases from their decay}

The fact that quantum gases decay away even as one wants to study them is certainly inconvenient.  Yet, this annoyance can also be turned to one's advantage.  In a paper entitled ``Coherence, correlations and collisions...'' Burt \emph{et al.}\ observed that Bose-Einstein condensates in their apparatus were decaying remarkably more slowly than non-degenerate gases \cite{burt97}. Examining closely the rate of loss of atoms from their gases, they found that three-body decay, losses that occur when a pair of atoms relax into a bound molecular state with the aid of a third collision partner, were diminished because the coherent matter-wave of a Bose-Einstein condensate contains fewer trios of atoms at short distances from each other than does an incoherent thermal gas.  The six-fold reduction of the zero-range third order correlation function, $g^{(3)}(0)$, is familiar also from the quantum optics of thermal and coherence sources of light.

Detailed studies of loss rates ascribed to two-, three-, and four-body processes have been used to reveal the short-range correlations that arise in a number of strongly interacting systems.  Suppressed losses revealed the ``fermionization'' of one-dimensional Bose gases as they enter the Tonks-Girardeau regime \cite{tolr04}.  Sharp increases and decreases in the three-body loss rate indicate the formation of three-body bound states due to strong two-body interactions \cite{kram06efimov}, as predicted by Efimov.  A reduced loss rate in two-state Fermi gases with strong interactions revealed the onset of local spin correlations and also the rapid formation of bound-state molecules in systems swept across Feshbach resonances \cite{jo09stoner}.

\subsection{Isolation vs.\ equilibrium}
\label{sec:isolation}

The lowest temperatures attained in cryogenic physics, in the $\mu$K range for bulk materials, are infernally high compared to the temperatures of quantum degenerate gases.  Thus, by necessity, atomic gases are kept extremely well isolated from their thermal environment by placing the gas within an ultrahigh vacuum environment, with pressures below $10^{-10}$ torr.  The residual background gas within the chamber, at thermal equilibrium with the chamber walls (say at room temperature), is so dilute that its effect is just to deplete the trapped gas slowly, since background-gas collisions typically impart more energy to an atom than the depth of trap in which it is contained.  Radiative coupling to the atomic gas is minimal since the atomic absorption spectrum is so sparse (atoms are a very low emissivity object).  This latter observation may have to be revisited for atoms and, particularly, for molecules with more dense and complex spectra.

This extreme isolation preserves the cold gas long enough for it to arrive at low temperature by evaporative cooling and to be manipulated and probed by the experimentalist.  However, the isolation also presents us with a challenge: without an external thermal bath with which to exchange energy, or an external reservoir with which to exchange particles, can the atomic materials be truly regarded as having reached equilibrium?  Thermodynamically, quantum gases should perhaps be described as a microcanonical ensemble; however, for a gas constantly subject to evaporative cooling, perhaps that picture is also inadequate.  In any case, one should exercise caution in importing findings from a grand-canonical quantum-statistical model, where systems are characterized by fixed chemical potential and temperature, rather than fixed energy and particle number.  The same concerns apply to quantum gases with a magnetic degree of freedom.  In condensed-matter studies of magnetism, the magnetic susceptibility is typically measured by applying a magnetic field to the sample, and measuring its resultant magnetization.  However, in quantum gases, the magnetization of the system can also be a conserved quantity given the absence of a ``spin reservoir.''  One way to connect the results of such models to the properties of gases trapped in inhomogeneous potentials is via the local density approximation, a method described above (Sec.\ \ref{sec:LDA}).

Aside from being limited by the presence of several conserved quantities, equilibration of quantum gases is also limited by the competition between the slowness of equilibration dynamics at low temperature and the finite lifetime or parasitic heating rates of the trapped gas.  As we demand increasingly precise information about the quantum gases in our laboratories, regarding, for example, the locations of transitions in a phase diagram or of subtle spatial features in a trapped gas, we require the atomic systems to be increasingly close to equilibrium, as highlighted by recent investigations of the effects of interactions on the Bose-Einstein condensation phase transition \cite{smit11} or of fluctuations in two-dimensional Bose-Hubbard systems \cite{hung10slowmass}.  The question of equilibration will continue to be important as cold-atom experiments make ever-deeper inroads into many-body quantum physics.

On the flip side, the persistence of non-equilibrium states in quantum gases may be turned into one of their most appealing features.  As discussed in the Chapter by Lamacraft and Moore, quantum gases offer us access to non-equilibrium dynamics in complex quantum systems that may be impossible to study otherwise.

\section{Conclusion}

Ultracold atomic physics provides an alternate approach to studying an increasing family of open problems in condensed-matter physics.  The experimental procedures are very different in the two fields, each being informed by a different intellectual legacy and by the different properties of the materials being studied.  But our understanding of the universal physics that is common to these fields is enriched by the complementary information and perspectives provided by the two research approaches.

It is tempting to distinguish atomic physics research as more of a ``materials by design'' rather than a ``materials by happenstance'' approach to materials science.  An idealistic view of this field posits that atomic physics experiments are a form of quantum simulation where one first identifies a many-body quantum physics model that one wants to understand better, and then one rigs up this model into an actual physical realization using cold atoms.  One can muse whether such simulations are to be regarded as experimental research per se, or whether they represent a form of ``experimental many-body quantum theory.''

However, the reality is that the field of cold atom physics is mostly driven by new experimental capabilities, rather than by the prior identification of a singular simulation goal. The systems selected for study are those that are just now becoming possible to study because of the development of a new technique for preparing the system or probing its properties.  Understanding these techniques is essential for knowing where the field is going next.  We hope the present Chapter, and the remaining works in this book, help convey this knowledge.

\section*{Acknowledgments}

We thank D.S.\ Jin and R.\ Hulet for permission to use figures from their works, J.\ Guzman for graphical assistance, and other members of
our research teams for their insights.  D.M.S-K acknowledges generous support from the NSF, AFOSR and DTRA.  J.H.T.\ acknowledges NSERC, AFOSR, and CIFAR. Both authors are supported by a grant from the ARO with funding from the DARPA OLE program.

\section*{References}

\bibliographystyle{elsarticle-num}



\end{document}